Chinonso I. Nwankwo[a,*], Weizhong Dai[b]

# Explicit RKF-Compact Scheme for Pricing Regime Switching American Options with Varying Time Step

[a] Department of Mathematics, Statistics, and Computer Science, University of Illinois at Chicago, Chicago, IL 60607, USA

[b] Department of Mathematics and Statistics, Louisiana Tech University, Ruston LA 71272, USA

[*] Corresponding author, nonsonwankwo@gmail.com
https://orcid.org/0000-0001-5526-1337

# Explicit RKF-Compact Scheme for Pricing Regime Switching American Options with Varying Time Step

## Abstract

In this research work, an explicit Runge-Kutta-Fehlberg (RKF) time integration with a fourth-order compact finite difference scheme in space and a high order analytical approximation of the optimal exercise boundary is employed for solving the regime-switching pricing model. In detail, we recast the free boundary problem into a system of nonlinear partial differential equations with a multi-fixed domain. We then introduce a transformation based on the square root function with a Lipschitz character from which a high order analytical approximation is obtained to compute the derivative of the optimal exercise boundary in each regime. We further compute the boundary values, asset option, and the option Greeks for each regime using fourth-order spatial discretization and adaptive time integration. In particular, the coupled assets options and option Greeks are estimated using Hermite interpolation with Newton basis. Finally, a numerical experiment is carried out with two- and four-regimes examples and results are compared with the existing methods. The results obtained from the numerical experiment show that the present method provides better performance in terms of computational speed and more accurate solutions with a large step size.

## 1. Mathematical Model

We present a mathematical model based on the American option put option with regime-switching. Let the option price $V_m(S, t)$ be written on the asset $S_t$ with strike price $K$, expiration time $T$, and $\tau = T - t$. Then, $V_m(S, \tau)$ satisfies the coupled free boundary value problem:

$$-\frac{\partial V_m(S,\tau)}{\partial \tau} + \frac{1}{2}\sigma^2{}_m S^2 \frac{\partial^2 V_m(S,\tau)}{\partial S^2} + r_m S \frac{\partial V_m(S,\tau)}{\partial S} - (r_m - q_{mm})V_m(S,\tau) + \sum_{l \neq m} q_{ml} V_l(S,\tau) = 0,$$

$$S > s_{f(m)}(\tau), \qquad (1a)$$

$$V_m(S,\tau) = K - S, \quad \text{for } S < s_{f(m)}(\tau). \qquad (1b)$$

Here, the initial and boundary conditions are given as:

$$V_m(S, 0) = max(K - S, 0), \qquad s_{f(m)}(0) = K; \qquad (1c)$$

$$V_m(s_{f(m)}, \tau) = K - s_{f(m)}(\tau), \quad V_m(\infty, \tau) = 0, \quad \frac{\partial}{\partial S}V_m(s_{f(m)}, \tau) = -1, \qquad (1d)$$

where $s_{f(m)}(\tau)$ is the optimal exercise boundary for the $m^{th}$ regime. $r_m$ and $\sigma_m$ are the interest rates and volatilities in each regime. The entries $q_{ml}$ of the generator matrix $Q_{I \times I}$ with $m, l = 1,2,\cdots, I$ satisfies the following relationship [23]:

$$q_{mm} = -\sum_{l \neq m} q_{ml}, \quad q_{ml} \geq 0, \quad for \ l \neq m, \quad l = 1,2,\cdots, I. \qquad (2)$$

Here, $q_{mm}$ represents the diagonal entries of $Q_{I \times I}$.

Several numerical methods have been proposed for solving the American option based on regime-switching [10, 22, 30, 4, 21, 8, 6, 26, 18, 19]. The commonly known method includes the penalty method [10, 20, 28], the method of line (MOL) [4, 21], the lattice method [8, 26], the fast Fourier transform [2, 19], and the front-fixing techniques [6]. Recently, the radial basis function generated finite difference method has also been implemented for solving the regime-switching model [16]. The numerical methods mentioned above provide up to second-order accurate solutions.

Moreover, some authors mentioned in their works of literature that computational challenges are encountered when solving the regime-switching model due to the coupled regime(s) and computing the optimal exercise boundary simultaneously with the options. Some of these challenges have led to imposing certain constraints on the model which include removing the optimal exercise boundary and introducing a penalty term [10, 20, 30], and treating the coupled regime explicitly when an implicit approach is implemented [6]. Some of these constraints reduce both the computational burden and accuracy of the numerical approximation which is substantially beyond the two-regimes example. Moreover, Chiarella et al. [4] mentioned that the coupled PDE model that describes option price based on regime-switching has not been fully explored and exploited. They further mentioned that the Greeks and optimal exercise boundary were not reported or unavailable in some of the existing works of literature. We also observed that the gamma option profile for each regime reported in some of the existing and current literature exhibits spurious oscillation beyond two-regimes example.

For the precise computation of the optimal exercise boundary, asset option, and sensitivities with high order accuracy in the American option pricing problem, we in [24] first derived a high order analytical approximation of the derivative of the optimal exercise boundary. With further implementation of the Runge-Kutta-Fehlberg time integration and compact finite difference method, highly accurate numerical solutions with high rates of convergence were achieved. In this work, we extend the approach in [24] to the regime-switching model. To this end, we first improve the boundary accuracy by deriving a highly accurate analytical approximation for computing the optimal exercise boundary in each regime. It enables us to present a fast, accurate, explicit, and high-order numerical scheme for solving the regime-switching model.

The rest of the paper is organized as follows. In section 2, we discuss the transformation methods involved in our method. In section 3, we discuss methods of interpolation and extrapolation for improving boundary accuracy. In section 4, we employ a compact scheme in the spatial discretization and Runge-Kutta-Fehlberg method for temporal discretization. In section 5, we perform a numerical experiment and compare the performance of our method with other existing methods. We then conclude the paper in section 6.

## 2. Transformations

### 2.1. Logarithmic Transformation

To compute the optimal exercise boundary simultaneously with the asset option and option Greeks, we first implement a front-fixing transformation [6,25,29] on the multi-variable domains as follows:

$$x_m = \ln\frac{S}{s_{f(m)}(\tau)} = \ln S - \ln s_{f(m)}(\tau), \qquad U_m(x_m, \tau) = V_m(S, \tau), \quad m = 1,2,\cdots,I. \tag{3}$$

The transformation in (3) enables us to fix the free boundary. Using this transformation and eliminating the first-order derivative by taking further derivatives, we then obtain a coupled system of nonlinear partial differential equations for each regime consisting of the asset option and its first derivative for each regime as follows:

$$\frac{\partial U_m}{\partial \tau} - \frac{1}{2}\sigma^2{}_m \frac{\partial^2 U_m}{\partial x_m^2} - \xi_m^\tau W_m + (r_m - q_{mm})U_m - \sum_{l\neq m} q_{ml}\, U_l = 0, \qquad x_m > 0; \tag{4a}$$

$$\frac{\partial W_m}{\partial \tau} - \frac{1}{2}\sigma^2{}_m \frac{\partial^2 W_m}{\partial x_m^2} - \xi_m^\tau \frac{\partial^2 U_m}{\partial x_m^2} + (r_m - q_{mm})W_m - \sum_{l\neq m} q_{ml}\, W_l = 0, \qquad x_m > 0, \tag{4b}$$

where $m = 1,2,\cdots,I$, $x_m \in [0,\infty)$. Here, $\xi_m^\tau = r_m - \sigma^2{}_m/2 + s'_{f(m)}(\tau)/s_{f(m)}(\tau)$. Furthermore, $W_m = \partial U_m/\partial x_m$ called the delta option. This Greek is an important hedging parameter and it will be ideal and reasonable to compute both the asset and delta options simultaneously with high order accuracy. Hence, it is not wasteful to include an additional PDE equation in (4b). A similar system of PDEs in (4) has been implemented in the work of Liao and Khaliq [17] and Dremkova and Ehrhardt [5] for solving the European option. The initial and boundary conditions for $U_m(x_m, \tau)$ and $W_m(x_m, \tau)$, are defined as:

$$U_m(x_m, 0) = \max(K - Ke^{x_m}, 0) = 0, \qquad x_m \geq 0, \qquad s_{f(m)}(0) = K; \tag{4c}$$

$$\lim_{x_m \to 0} U_m(x_m, \tau) = K - s_{f(m)}(\tau), \qquad \lim_{x_m \to \infty} U_m(x_m, \tau) = 0; \tag{4d}$$

$$\lim_{x_m \to 0} W_m(x_m, \tau) = -s_{f(m)}(\tau), \qquad \lim_{x_m \to \infty} W_m(x_m, \tau) = 0; \tag{4e}$$

$$U_m(x_m, 0) = 0, \qquad W_m(x_m, 0) = 0, \ x_m \geq 0. \tag{4f}$$

### 2.2. Square Root Transformation

In this subsection, a square root transformation [11,12,13,14,15,24] is employed for approximating the optimal exercise boundaries in the regime-switching model. Kim et al. [11] claimed that with this transformation that has Lipschitz surface, the degeneracy that occurs in the method of Wu and Kwok [29] near the optimal exercise boundary can be avoided. It is worth noting that a good approximation of the optimal exercise boundary influences the accuracy of the option price, especially, when a high order numerical scheme is implemented [20,24]. The advantage of this transformation with Lipschitz's character near the optimal exercise boundary lies in the accurate computation of the latter with much simplicity and no iteration. Here, we implement this square root transformation to the regime-switching model. To this end, we first write the square root transformation as follows:

$$Q_m(x_m, \tau) = \sqrt{U_m(x_m, \tau) - K + e^{x_m} s_{f(m)}(\tau)} \qquad \text{or} \qquad U_m(x_m, \tau) = Q_m^2(x_m, \tau) + K - e^{x_m} s_{f(m)}(\tau). \tag{5}$$

Here,

$$Q_m(x_m, \tau) \begin{cases} = 0, & x_m \in [\ln s_{f(m)}(\infty) - \ln s_{f(m)}(0)], \\ > 0, & x_m \in (0, \infty). \end{cases} \tag{6}$$

To compute the optimal exercise boundary for each regime, we first evaluate the derivative of the square root function at the fixed boundary point up to the third-order derivative. The first-order derivative of the square root function is computed as follows:

$$U_m(0, \tau) = K - s_{f(m)}(\tau), \tag{7a}$$

$$(U_m)_{x_m}(x_m, \tau) = 2Q_m(x_m, \tau)(Q_m)_{x_m}(x_m, \tau) - e^{x_m} s_{f(m)}(\tau), \qquad (U_m)_{x_m}(0, \tau) = -s_{f(m)}(\tau); \tag{7b}$$

$$(U_m)_{x_m x_m}(x_m, \tau) = 2Q_m(x_m, \tau)(Q_m)_{x_m x_m}(x_m, \tau) + 2\left((Q_m)_{x_m}(x_m, \tau)\right)^2 - e^{x_m} s_{f(m)}(\tau); \tag{7c}$$

$$(U_m)_{x_m x_m}(0,\tau) = 2\left((Q_m)_{x_m}(0,\tau)\right)^2 - s_{f(m)}(\tau). \tag{7d}$$

Following the approach in the work of Goodman and Ostrov [7] for computing $(U_m)_\tau(x_m = 0, \tau)$, we differentiate (7a) at the optimal exercise boundary with respect to $\tau$ and obtain

$$(U_m)_\tau(0,\tau) = -s'_{f(m)}(\tau). \tag{7e}$$

Substituting (7) into (4a) when $x_m = 0$, we obtain

$$-s'_{f(m)}(\tau) - \frac{\sigma_m^2}{2}\left[2\left((Q_m)_{x_m}(0,\tau)\right)^2 - e^{x_m}s_{f(m)}(\tau)\right] + \xi_m^\tau s_{f(m)}(\tau) + (r_m - q_{mm})\left(K - s_{f(m)}(\tau)\right)$$
$$- \sum_{l \neq m} q_{ml}\, U_l(x_l | x_m = 0, \tau) = 0. \tag{8}$$

From (8), the first derivative of the square root function at $x_m = 0$ is given as follows:

$$(Q_m)_{x_m}(0,\tau) = Q'_m(0,\tau) = \frac{\sqrt{(r_m - q_{mm})K + q_{mm}s_{f(m)}(\tau) - \sum_{l \neq m} q_{ml}\, U_l(x_l | x_m = 0, \tau)}}{\sigma_m}. \tag{9}$$

We then proceed to compute the second derivative of the square root function for each regime as follows:

$$(U_m)_{x_m \tau}(0,\tau) = -s'_{f(m)}(\tau), \tag{10a}$$

$$(U_m)_{x_m x_m x_m}(x_m, \tau)$$
$$= 2Q_m(x_m, \tau)(Q_m)_{x_m x_m x_m}(x_m, \tau) + 6(Q_m)_{x_m}(x_m, \tau)(Q_m)_{x_m x_m}(x_m, \tau)$$
$$- e^{x_m}s_{f(m)}(\tau), \tag{10b}$$

$$(U_m)_{x_m x_m x_m}(0,\tau) = 6(Q_m)_{x_m}(0,\tau)(Q_m)_{x_m x_m}(0,\tau) - s_{f(m)}(\tau). \tag{10c}$$

Substituting (10) into (4b) and simplifying further when $x_m = 0$, we obtain

$$-s'_{f(m)}(\tau) - \frac{\sigma_m^2}{2}\left[6(Q_m)_{x_m}(0,\tau)(Q_m)_{x_m x_m}(0,\tau) - s_{f(m)}(\tau)\right] + \xi_m^\tau\left[2\left((Q_m)_{x_m}\right)^2 - s_{f(m)}(\tau)\right]$$
$$- (r_m - q_{mm})s_{f(m)}(\tau) - \sum_{l \neq m} q_{ml}\, U'_l(x_l | x_m = 0, \tau) = 0, \tag{11}$$

$$(Q_m)_{x_m x_m}(0,\tau) = Q''_m(0,\tau)$$
$$= -\frac{2\xi_m^\tau Q'_m(0,\tau)}{3\sigma_m^2} + \frac{q_{mm}s_{f(m)}(\tau)}{3\sigma_m^2 Q'_m(0,\tau)} - \frac{\sum_{l \neq m} q_{ml}\, U'_l(x_l | x_m = 0, \tau)}{3\sigma_m^2 Q'_m(0,\tau)}. \tag{12}$$

Finally, we compute the third derivatives of the square root function at $x_m = 0$ as follows:

$$(U_m)_{x_m x_m x_m x_m}(x_m, \tau)$$
$$= 2Q_m(x_m, \tau)(Q_m)_{x_m x_m x_m x_m}(x_m, \tau) + 8(Q_m)_{x_m}(x_m, \tau)(Q_m)_{x_m x_m x_m}(x_m, \tau)$$
$$+ 6\left((Q_m)_{x_m x_m}(x_m, \tau)\right)^2 - e^{x_m} s_{f(m)}(\tau), \tag{13a}$$

$$(U_m)_{x_m x_m x_m x_m}(0, \tau) = 8(Q_m)_{x_m}(0, \tau)(Q_m)_{x_m x_m x_m}(0, \tau) + 6\left((Q_m)_{x_m x_m}(0, \tau)\right)^2 - s_{f(m)}(\tau). \tag{13b}$$

For $(U_m)_{x_m x_m \tau}(0, \tau)$, if $x_m \to 0^+$, we need to differentiate (7d) with respect to the time variable as follow:

$$(U_m)_{x_m x_m \tau}(0, \tau) = \frac{\partial}{\partial \tau}\left(\frac{2(r_m - q_{mm})K + 2q_{mm}s_{f(m)}(\tau) - 2\sum_{l \neq m} q_{ml} U_l(x_l | x_m = 0, \tau)}{\sigma_m^2} - s_{f(m)}(\tau)\right) \tag{14a}$$

It will be challenging to differentiate (14a) due to the coupled regime. Here, for simplicity, we consider instead $x_m \to 0^-$ and compute as follows:

$$(U_m)_{x_m x_m \tau}(0, \tau) = -s'_{f(m)}(\tau), \tag{14b}$$

Furthermore, differentiating (4b) with respect to $x_m$ and substituting (14) when $x_m = 0$, we obtain

$$-s'_{f(m)}(\tau) - \frac{\sigma_m^2}{2}\left[8(Q_m)_{x_m}(0, \tau)(Q_m)_{x_m x_m x_m}(0, \tau) + 6\left((Q_m)_{x_m x_m}(0, \tau)\right)^2 - s_{f(m)}(\tau)\right]$$
$$+ \xi_m^\tau [6(Q_m)_{x_m}(0, \tau)(Q_m)_{x_m x_m}(0, \tau) - s_{f(m)}(\tau)]$$
$$+ (r_m - q_{mm})\left[2\left((Q_m)_{x_m}(0, \tau)\right)^2 - s_{f(m)}(\tau)\right] - \sum_{l \neq m} q_{ml} U_l''(x_l | x_m = 0, \tau) = 0. \tag{15a}$$

Simplifying further, we obtain the third derivative of the square root functions at $x_m = 0$ as follows

$$(Q_m)_{x_m x_m x_m}(0, \tau) = Q_m'''(0, \tau)$$
$$= \frac{2(\xi_m^\tau)^2 Q_m'(0, \tau)}{3\sigma_m^4} - \xi_m^\tau \left[\frac{q_{mm} s_{f(m)}(\tau)}{6\sigma_m^4 Q_m'(0, \tau)} - \frac{\sum_{l \neq m} q_{ml} U_l'(x_l | x_m = 0, \tau)}{6\sigma_m^4 Q_m'(0, \tau)}\right]$$
$$- \frac{3}{4Q_m'(0, \tau)}\left[\left(\frac{q_{mm} s_{f(m)}(\tau)}{3\sigma_m^2 Q_m'(0, \tau)}\right)^2 + \left(\frac{\sum_{l \neq m} q_{ml} U_l'(x_l | x_m = 0, \tau)}{3\sigma_m^2 Q_m'(0, \tau)}\right)^2\right]$$
$$- \frac{2q_{mm} s_{f(m)}(\tau) \sum_{l \neq m} q_{ml} U_l'(x_l | x_m = 0, \tau)}{[3\sigma_m^2 Q_m'(0, \tau)]^2} + \frac{Q_m'(0, \tau)(r_m - q_{mm})}{2\sigma_m^2} + \frac{q_{mm} s_{f(m)}(\tau)}{4\sigma_m^2 Q_m'(0, \tau)}$$
$$- \frac{\sum_{l \neq m} q_{ml} U_l''(x_l | x_m = 0, \tau)}{4\sigma_m^2 Q_m'(0, \tau)}. \tag{15b}$$

To approximate $U_l(x_l | x_m = 0, \tau), U_l'(x_l | x_m = 0, \tau)$, and $U_l''(x_l | x_m = 0, \tau)$ in (9), (12), and (15b) we consider the following relationship

$$x_l = 0 - \ln\frac{s_{f(l)}(\tau_n)}{s_{f(m)}(\tau_n)}. \tag{16}$$

If $x_l \leq 0$, then $U_l(x_l|x_m = 0, \tau) = K - e^{(x_l)_{j^*}} s_{f(l)}$, $U_l'(x_l|x_m = 0, \tau) = -e^{(x_l)_{j^*}} s_{f(l)}$, and $U_l''(x_l|x_m = 0, \tau) = -e^{(x_l)_{j^*}} s_{f(l)}$. However, if $x_l > 0$, we need an interpolation method to approximate $U_l(x_l|x_m = 0, \tau)$, $U_l'(x_l|x_m = 0, \tau)$, and $U_l''(x_l|x_m = 0, \tau)$. Here, we use the quintic Hermite interpolation with Newton basis to approximate $U_l''(x_l|x_m = 0, \tau)$ at least with a third-order accuracy. Let $j^* \in [j, j+2]$ be the point in the interval of $l^{th}$ regime that we need to approximate $U_l(x_l|x_m = 0, \tau)$, $U_l'(x_l|x_m = 0, \tau)$, and $U_l''(x_l|x_m = 0, \tau)$, then the quintic Hermite function and its first and second derivatives, are presented as follows:

$$U_l(x_l|x_m = 0, \tau)$$
$$\approx \alpha_0 + \alpha_1[(x_l)_{j^*} - (x_l)_j] + \alpha_2[(x_l)_{j^*} - (x_l)_j]^2 + \alpha_3[(x_l)_{j^*} - (x_l)_j]^2[(x_l)_{j^*} - (x_l)_{j+1}]$$
$$+ \alpha_4[(x_l)_{j^*} - (x_l)_j]^2[(x_l)_{j^*} - (x_l)_{j+1}]^2$$
$$+ \alpha_5[(x_l)_{j^*} - (x_l)_j]^2[(x_l)_{j^*} - (x_l)_{j+1}]^2[(x_l)_{j^*} - (x_l)_{j+2}], \tag{17a}$$

$$U_l'(x_l|x_m = 0, \tau)$$
$$\approx \alpha_1 + 2\alpha_2[(x_l)_{j^*} - (x_l)_j] + 2\alpha_3[(x_l)_{j^*} - (x_l)_j][(x_l)_{j^*} - (x_l)_{j+1}] + \alpha_3[(x_l)_{j^*} - (x_l)_j]^2$$
$$+ 2\alpha_4[(x_l)_{j^*} - (x_l)_j][(x_l)_{j^*} - (x_l)_{j+1}]^2 + 2\alpha_4[(x_l)_{j^*} - (x_l)_{j+1}][(x_l)_{j^*} - (x_l)_j]^2$$
$$+ 2\alpha_5[(x_l)_{j^*} - (x_l)_j][(x_l)_{j^*} - (x_l)_{j+1}]^2[(x_l)_{j^*} - (x_l)_{j+2}]$$
$$+ 2\alpha_5[(x_l)_{j^*} - (x_l)_j]^2[(x_l)_{j^*} - (x_l)_{j+1}][(x_l)_{j^*} - (x_l)_{j+2}]$$
$$+ \alpha_5[(x_l)_{j^*} - (x_l)_j]^2[(x_l)_{j^*} - (x_l)_{j+1}]^2, \tag{17b}$$

$$U_l''(x_l|x_m = 0, \tau)$$
$$\approx 2\alpha_2 + 4\alpha_3[(x_l)_{j^*} - (x_l)_j] + 2\alpha_3[(x_l)_{j^*} - (x_l)_{j+1}] + 2\alpha_4[(x_l)_{j^*} - (x_l)_{j+1}]^2$$
$$+ 2\alpha_4[(x_l)_{j^*} - (x_l)_j]^2 + 2\alpha_4[(x_l)_{j^*} - (x_l)_{j+1}]^2 + 8\alpha_4[(x_l)_{j^*} - (x_l)_j][(x_l)_{j^*} - (x_l)_{j+1}]$$
$$+ 4\alpha_5[(x_l)_{j^*} - (x_l)_{j+1}]^2[(x_l)_{j^*} - (x_l)_j] + 2\alpha_5[(x_l)_{j^*} - (x_l)_{j+1}]^2[(x_l)_{j^*} - (x_l)_{j+2}]$$
$$+ 8\alpha_5[(x_l)_{j^*} - (x_l)_j][(x_l)_{j^*} - (x_l)_{j+1}]^2[(x_l)_{j^*} - (x_l)_{j+2}]$$
$$+ 4\alpha_5[(x_l)_{j^*} - (x_l)_j]^2[(x_l)_{j^*} - (x_l)_{j+1}] + 2\alpha_5[(x_l)_{j^*} - (x_l)_j]^2[(x_l)_{j^*} - (x_l)_{j+2}], \tag{17c}$$

where

$$\alpha_0 = (u_l)_{j|0}^n, \quad \alpha_1 = (w_l)_{j|0}^n, \quad \alpha_2 = \frac{1}{h}\left(\frac{(u_l)_{j+1|0}^n - (u_l)_{j|0}^n}{h} - (w)_{j|0}^n\right); \tag{17d}$$

$$\alpha_3 = \frac{1}{h}\left[\frac{1}{h}\left((w)_{j+1|0}^n - \frac{(u_l)_{j+1|0}^n - (u_l)_{j|0}^n}{h}\right) - \alpha_2\right], \tag{17e}$$

$$\alpha_4 = \frac{1}{2h}\left(\frac{1}{2h}\left[\frac{1}{h}\left(\frac{(u_l)_{j+2|0}^n - (u_l)_{j+1|0}^n}{h} - (w)_{j+1|0}^n\right) - \frac{1}{h}\left((w)_{j+1|0}^n - \frac{(u_l)_{j+1|0}^n - (u_l)_{j|0}^n}{h}\right)\right] - \alpha_3\right), \tag{17f}$$

$$\alpha_5 = \frac{1}{2h}\left[\left(\frac{1}{h}\left[\frac{1}{h}\left((w)_{j+2|0}^n - \frac{(u_l)_{j+2|0}^n - (u_l)_{j+1|0}^n}{h}\right) - \frac{1}{h}\left(\frac{(u_l)_{j+2|0}^n - (u_l)_{j+1|0}^n}{h} - (w)_{j+1|0}^n\right)\right]\right.\right.$$
$$\left.\left. - \frac{1}{2h}\left[\frac{1}{h}\left(\frac{(u_l)_{j+2|0}^n - (u_l)_{j+1|0}^n}{h} - (w)_{j+1|0}^n\right) - \frac{1}{h}\left((w)_{j+1|0}^n - \frac{(u_l)_{j+1|0}^n - (u_l)_{j|0}^n}{h}\right)\right]\right)\right.$$
$$\left. - \alpha_4\right]. \tag{17g}$$

We now derive a high order analytical approximation of the derivative of the optimal exercise boundary for each regime by employing extrapolated Taylor expansion of the intermediate function $Q_m(\bar{x}_m, \tau)$ at $x_m = 0$. Here, $\bar{x}_m$ denotes arbitrary point very close to the optimal exercise boundary for each regime's interval. We first obtain the following lemma.

**Lemma.** Assume $Q_m(x,\tau) \in C^6[0, 3\bar{x}_m]$, then it holds

$$a_0 Q_m(0,\tau) + a_1 Q_m(\bar{x}_m, \tau) + a_2 Q_m(2\bar{x}_m, \tau) + a_3 Q_m(3\bar{x}_m, \tau)$$
$$= b_1 \bar{x}_m Q'_m(0,\tau) + b_2 \bar{x}_m^2 Q''_m(0,\tau) + b_3 \bar{x}_m^3 Q'''_m(0,\tau) + O(\bar{x}_m^6), \tag{18}$$

with

$$a_0 = -\frac{175}{4}, \quad a_1 = 81, \quad a_2 = -\frac{81}{8}, \quad a_3 = 1, \quad b_1 = \frac{255}{4}, \quad b_2 = \frac{99}{4}, \quad b_3 = \frac{9}{2}. \tag{19}$$

**Proof.** We refer the readers to the work of Chinonso and Dai [24].

Substituting (9), (12), (15b), and (17) into (18b), we then obtain a high order analytical approximation for each regime in the form of a quadratic equation as follows:

$$(d_m)_1 \left(s'_{f(m)}(\tau)\right)^2 + (d_m)_2 s'_{f(m)}(\tau) + (d_m)_3 = 0, \quad m = 1, 2, \cdots I, \tag{20}$$

with

$$(d_m)_1 = \frac{2 b_3 Q'_m(0,\tau) \bar{x}_m^3}{3 \sigma_m^4 s_{f(m)}^2(\tau)}, \tag{21a}$$

$$(d_m)_2 = \frac{4 b_3 v_m Q'_m(0,\tau) \bar{x}_m^3}{3 \sigma_m^4 s_{f(m)}(\tau)} - \frac{2 b_2 Q'_m(0,\tau) \bar{x}_m^2}{3 \sigma_m^2 s_{f(m)}(\tau)} - \frac{b_3 \bar{x}_m^3 q_{mm}}{6 \sigma_m^4 Q'_m(0,\tau)} + \frac{b_3 \bar{x}_m^3 \sum_{l \neq m} q_{ml} U'_l(x_l | x_m = 0, \tau)}{6 \sigma_m^4 Q'_m(0,\tau) s_{f(m)}(\tau)}, \tag{21b}$$

$$(d_m)_3 = \frac{2v_m^2 Q_m'(0,\tau)\bar{x}_m^3}{3\sigma_m^4} - v_m \left[\frac{2b_2 Q_m'(0,\tau)\bar{x}_m^2}{3\sigma_m^2} + \frac{b_3 q_{mm} s_{f(m)}(\tau)\bar{x}_m^3}{6\sigma_m^4 Q_m'(0,\tau)} - \frac{b_3 \bar{x}_m^3 \sum_{l\neq m} q_{ml} U_l'(x_l|x_m=0,\tau)}{6\sigma_m^4 Q_m'(0,\tau)}\right]$$

$$- \frac{3b_3 \bar{x}_m^3}{4Q_m'(0,\tau)}\left[\left(\frac{q_{mm} s_{f(m)}(\tau)}{3\sigma_m^2 Q_m'(0,\tau)}\right)^2 + \left(\frac{\sum_{l\neq m} q_{ml} U_l'(x_l|x_m=0,\tau)}{3\sigma_m^2 Q_m'(0,\tau)}\right)^2\right.$$

$$\left. - \frac{2q_{mm} s_{f(m)}(\tau)\sum_{l\neq m} q_{ml} U_l'(x_l|x_m=0,\tau)}{[3\sigma_m^2 Q_m'(0,\tau)]^2}\right]$$

$$+ \frac{Q_m'(0,\tau) b_3 (r_m - q_{mm})\bar{x}_m^3}{2\sigma_m}$$

$$+ \frac{b_2 \bar{x}_m^2}{3\sigma_m^2 Q_m'(0,\tau)}\left[q_{mm} s_{f(m)}(\tau) - \sum_{l\neq m} q_{ml} U_l'(x_l|x_m=0,\tau)\right]$$

$$+ \frac{b_3 \bar{x}_m^3}{4\sigma_m^2 Q_m'(0,\tau)}\left[q_{mm} s_{f(m)}(\tau) - \sum_{l\neq m} q_{ml} U_l''(x_l|x_m=0,\tau)\right] - a_1 Q_m(\bar{x}_m,\tau)$$

$$- a_2 Q_m(2\bar{x}_m,\tau) - a_3 Q_m(3\bar{x}_m,\tau). \tag{21c}$$

Note that $v_m = r_m - \sigma_m^2/2$. The derivative of the optimal exercise boundary for each regime is then computed as follows:

$$s'_{f(m)}(\tau) = \frac{-(d_m)_2 - \sqrt{[(d_m)_2]^2 - 4(d_m)_1 (d_m)_3}}{2(d_m)_1}, \quad m = 1,2,\cdots I. \tag{22}$$

Without the above extrapolation in the lemma, only a second order accurate solution should be expected as seen in [24] even if the fourth-order compact method is employed for Eq. (4). With the method of extrapolation, we ensure that a high order approximation in space is maintained at the boundary for each regime. To incorporate the higher-order approximation of the optimal exercise boundary, we will employ the compact finite difference method for spatial discretization, cubic Hermite with Newton Basis for interpolation, and high order adaptive Runge-Kutta method for temporal discretization. This is detailed in the following section.

## 3. Numerical Method

For each regime, the discretized system of PDEs is solved in a uniform space grid and non-uniform adaptive time grid $[0,\infty) \times [0\ T]$. Following the approach in [9,27], the infinite space domain is replaced with the truncated far boundary $x_{fb}$. Letting $i$ and $j$ represent the node points in the $l^{th}$ and $m^{th}$ regimes' intervals, respectively, and M represents the numbers of grid points. Then we have

$$(x_m)_i = ih, \quad (x_l)_j = jh, \quad h = \frac{x_{fb}}{M}, \quad i,j \in [0,M]. \tag{23}$$

The numerical approximations of the asset and delta options and the optimal exercise boundary for each regime are denoted as $(u_m)_i^n$, $(w_m)_i^n$, and $s_{f(m)}^n$, respectively.

## 3.1. Compact Finite Difference Scheme

For each regime, we implement a compact scheme for discretization of the system of a partial differential equation that consists of the asset option and delta sensitivity. In the interior points, we use the compact scheme discretization as follows:

$$f''(x_{i-1}) + 10f''(x_i) + f''(x_{i+1}) = \frac{12}{h^2}[f(x_{i-1}) - 2f(x_i) + f(x_{i+1})] + O(h^4). \tag{24}$$

For the exterior points, we implement a one-sided fourth order scheme as follows [31, 1]:

$$14f''(x_1) - 5f''(x_2) + 4f''(x_3) - f''(x_4) = \frac{12}{h^2}[f(x_0) - 2f(x_1) + f(x_2)] + O(h^4). \tag{25a}$$

$$14f''(x_{M-1}) - 5f''(x_{M-2}) + 4f''(x_{M-3}) - f''(x_{M-4})$$
$$= \frac{12}{h^2}[f(x_{M-2}) - 2f(x_{M-1}) + f(x_M)] + O(h^4). \tag{25b}$$

We then present (24-25) in matrix-vector form as follows:

$$\boldsymbol{u}_m'' = B^{-1}[A\boldsymbol{u}_m + (\boldsymbol{f}_m)_u], \qquad \boldsymbol{w}_m'' = B^{-1}[A\boldsymbol{w}_m + (\boldsymbol{f}_m)_w]. \tag{26}$$

with

$$A = \frac{12}{h^2}\begin{bmatrix} -2 & 1 & 0 & \cdots & & & & 0 \\ 1 & -2 & 1 & & & & & \vdots \\ & 1 & -2 & 1 & & & & \\ & & 1 & -2 & 1 & & & \\ 0 & & & \ddots & \ddots & \ddots & & 0 \\ & & & & 1 & -2 & 1 & \\ \vdots & & & & & 1 & -2 & 1 \\ 0 & & & & \cdots & 0 & 1 & -2 \end{bmatrix}_{M-1 \times M-1},$$

$$B = \begin{bmatrix} 14 & -5 & 4 & -1 & 0 & \cdots & 0 \\ 1 & 10 & 1 & & & & \vdots \\ & 1 & 10 & 1 & & & \\ & & 1 & 10 & 1 & & \\ 0 & & & \ddots & \ddots & \ddots & 0 \\ \vdots & & & & 1 & 10 & 1 \\ 0 & \cdots & 0 & -1 & 4 & -5 & 14 \end{bmatrix}_{M-1 \times M-1},$$

$$(\boldsymbol{f}_m)_u = \frac{12}{h^2}\begin{bmatrix}(u_m)_0 \\ 0 \\ 0 \\ \vdots \\ 0 \\ (u_m)_M = 0\end{bmatrix}_{M-1 \times 1} \quad (\boldsymbol{f}_m)_w = \frac{12}{h^2}\begin{bmatrix}(w_m)_0 \\ 0 \\ 0 \\ \vdots \\ 0 \\ (w_m)_M = 0\end{bmatrix}_{M-1 \times 1}. \tag{27}$$

Substituting (26) in (4), we discretize in the spatial direction and recast (4) in the form of a system of ordinary differential equations as follows:

$$\frac{\partial \boldsymbol{u}_m}{\partial \tau} = \boldsymbol{g}_1(\boldsymbol{u}_m, \boldsymbol{w}_m, \boldsymbol{u}_l), \qquad \frac{\partial \boldsymbol{w}_m}{\partial \tau} = \boldsymbol{g}_2(\boldsymbol{w}_m, \boldsymbol{u}_m, \boldsymbol{w}_l), \tag{28}$$

where

$$\boldsymbol{g}_1 = \frac{\sigma_m^2}{2} B^{-1}[A\boldsymbol{u}_m + (\boldsymbol{f}_m)_u] + \xi_\tau^m \boldsymbol{w}_m - (r_m - q_{mm})\boldsymbol{u}_m + \sum_{l \neq m} q_{ml}\, \boldsymbol{u}_l, \tag{29a}$$

$$\boldsymbol{g}_2 = \frac{\sigma_m^2}{2} B^{-1}[A\boldsymbol{w}_m + (\boldsymbol{f}_m)_w] + \xi_\tau^m B^{-1}[A\boldsymbol{u}_m + (\boldsymbol{f}_m)_u] - (r_m - q_{mm})\boldsymbol{w}_m + \sum_{l \neq m} q_{ml}\, \boldsymbol{w}_l. \tag{29b}$$

For $x_l < 0$ and $x_l \in (0, x_{fb})$, $u_l(x_l|x_m, \tau)$ and $w_l(x_l|x_m, \tau)$ are approximated using the procedure as described in (18), subsection 3.1. However, for fast computation, we use cubic Hermite interpolation with a Newton basis. For $x_l > x_{fb}$, we set $U_l(x_l|x_m = 0, \tau) = U_l'(x_l|x_m = 0, \tau) = U_l''(x_l|x_m = 0, \tau) = 0$.

### 3.2. Embedded RKF Time Integration

In this subsection, we present an explicit approach for solving (20) and (29) using the adaptive RKF method by first recasting (20) and (29) as follows:

$$\frac{\partial s_{f(m)}^n}{\partial \tau} = g\big(s_{f(m)}^n, u_{\bar{x}}^n\big) = \frac{-(d_m^n)_2 - \sqrt{[(d_m^n)_2]^2 - 4(d_m^n)_1(d_m^n)_3}}{2(d_m^n)_1}, \tag{30a}$$

with

$$\frac{\partial \boldsymbol{u}_m^n}{\partial \tau} = \frac{\sigma_m^2}{2} B^{-1}[A\boldsymbol{u}_m^n + (\boldsymbol{f}_m^n)_u] + \xi_n^m \boldsymbol{w}_m^n - (r_m - q_{mm})\boldsymbol{u}_m^n + \sum_{l \neq m} q_{ml}\, \boldsymbol{u}_l^n, \tag{30b}$$

$$\frac{\partial w_m^n}{\partial \tau} = \frac{\sigma_m^2}{2} B^{-1}[A w_m^n + (f_m^n)_w] + \xi_n^m B^{-1}[A u_m^n + (f_m^n)_u] - (r_m - q_{mm}) w_m^n + \sum_{l \neq m} q_{ml} w_l^n. \quad (30c)$$

Here, we use the RKF method based on the coefficient of Cash and Karp [3]. For brevity, we only present the general form of the RKF method for the temporal discretization of (30a) and (30b). For the approximation of the optimal exercise boundary for each regime, the fifth-order Runge-Kutta method

$$s_{f(m)}^{n+1} = s_{f(m)}^n + \left( \frac{37}{378} R_{S_{f(m)}}^1 + \frac{250}{621} R_{S_{f(m)}}^3 + \frac{125}{594} R_{S_{f(m)}}^4 + \frac{512}{1771} R_{S_{f(m)}}^5 \right), \quad (31a)$$

is computed simultaneously with the fourth-order Runge-Kutta method

$$\bar{s}_{f(m)}^{n+1} = s_{f(m)}^n + \left( \frac{2825}{27648} R_{S_{f(m)}}^1 + \frac{18575}{48384} R_{S_{f(m)}}^3 + \frac{13525}{55296} R_{S_{f(m)}}^4 + \frac{277}{14336} R_{S_{f(m)}}^5 + \frac{1}{4} R_{S_{f(m)}}^6 \right), \quad (31b)$$

and the error estimate $e_{S_f} = \left| s_{f(m)}^{n+1} - \bar{s}_{f(m)}^{n+1} \right| < \varepsilon$. Here,

$$R_{S_{f(m)}}^1 = g\left(s_{f(m)}^n, u_{\bar{x}_m}^n\right) k, \quad R_{S_{f(m)}}^2 = g\left(s_{f(m)}^n + \frac{1}{5} R_{S_{f(m)}}^1, u_{\bar{x}_m}^n\right) k; \quad (32d)$$

$$R_{S_{f(m)}}^3 = g\left(s_{f(m)}^n + \frac{3}{40} R_{S_{f(m)}}^1 + \frac{9}{40} R_{S_{f(m)}}^2, u_{\bar{x}_m}^n\right) k, \quad (32e)$$

$$R_{S_{f(m)}}^4 = g\left(s_{f(m)}^n + \frac{3}{10} R_{S_{f(m)}}^1 - \frac{9}{10} R_{S_{f(m)}}^2 + \frac{6}{5} R_{S_{f(m)}}^3, u_{\bar{x}_m}^n\right) k, \quad (32f)$$

$$R_{S_{f(m)}}^5 = g\left(s_{f(m)}^n - \frac{11}{54} R_{S_{f(m)}}^1 + \frac{5}{2} R_{S_{f(m)}}^2 - \frac{70}{27} R_{S_{f(m)}}^3 + \frac{35}{27} R_{S_{f(m)}}^4, u_{\bar{x}_m}^n\right) k, \quad (32g)$$

$$R_{S_f}^6 = g\left(s_{f(m)}^n + \frac{1631}{55296} R_{S_{f(m)}}^1 + \frac{175}{512} R_{S_{f(m)}}^2 + \frac{575}{13824} R_{S_{f(m)}}^3 + \frac{44275}{110592} R_{S_{f(m)}}^4 + \frac{253}{4096} R_{S_{f(m)}}^5, u_{\bar{x}_m}^n\right) k. \quad (32h)$$

Here, we choose $\bar{x} = 4h$ in our numerical experiment. The notation $k$ represents the time step.

**Remark 1**. For each evaluation of $R_{S_{f(m)}}^j$ and

$$s_{f(m)}^{n(j)} = s_{f(m)}^n + \sum_{i=1}^{j} a_i R_{S_{f(m)}}^i,$$

$U_l(x_l | x_m = 0, \tau), U_l'(x_l | x_m = 0, \tau)$, and $U_l''(x_l | x_m = 0, \tau)$ are re-evaluated based on section 3.1. This is because whenever the value of the optimal exercise boundary changes for each regime, it is likely that the location of the $U_l(x_l | x_m = 0, \tau), U_l'(x_l | x_m = 0, \tau)$, and $U_l''(x_l | x_m = 0, \tau)$ will change based on the relationship in (16).

To approximate the asset option for each regime, the fifth-order Runge-Kutta method

$$u_m^{n+1} = u_m^n + \left(\frac{37}{378}L_{u_m}^1 + \frac{250}{621}L_{u_m}^3 + \frac{125}{594}L_{u_m}^4 + \frac{512}{1771}L_{u_m}^6\right), \qquad (33a)$$

is computed simultaneously with the fourth-order Runge-Kutta method

$$\bar{u}_m^{n+1} = u_m^n + \left(\frac{2825}{27648}L_{u_m}^1 + \frac{18575}{48384}L_{u_m}^3 + \frac{13525}{55296}L_{u_m}^4 + \frac{277}{14336}L_{u_m}^5 + \frac{1}{4}L_{u_m}^6\right), \qquad (33b)$$

respectively, and the error estimated as

$$e_u = \max_{1 \leq m \leq I}\left\|u_m^{n+1} - \bar{u}_m^{n+1}\right\|_\infty < \varepsilon, \qquad (33c)$$

where

$$L_{u_m}^1 = g(u_m^n, w_m^n, u_l^n)k, \qquad L_{u_m}^2 = g\left(u_m^n + \frac{1}{5}L_{u_m}^1, w_m^n + \frac{1}{5}L_{w_m}^1, u_l^n\right)k; \qquad (34a)$$

$$L_{u_m}^3 = g\left(u_m^n + \frac{3}{40}L_{u_m}^1 + \frac{9}{40}L_{u_m}^2, w_m^n + \frac{3}{40}L_{w_m}^1 + \frac{9}{40}L_{w_m}^2, u_l^n\right)k, \qquad (34b)$$

$$L_{u_m}^4 = g\left(u_m^n + \frac{3}{10}L_{u_m}^1 - \frac{9}{10}L_{u_m}^2 + \frac{6}{5}L_{u_m}^3, w_m^n + \frac{3}{10}L_{w_m}^1 - \frac{9}{10}L_{w_m}^2 + \frac{6}{5}L_{w_m}^3, u_l^n\right)k, \qquad (34c)$$

$$L_{u_m}^5 = g\left(u_m^n - \frac{11}{54}L_{u_m}^1 + \frac{5}{2}L_{u_m}^2 - \frac{70}{27}L_{u_m}^3 + \frac{35}{27}L_{u_m}^4, w_m^n - \frac{11}{54}L_{w_m}^1 + \frac{5}{2}L_{w_m}^2 - \frac{70}{27}L_{w_m}^3 \right.$$
$$\left. + \frac{35}{27}L_{w_m}^4, u_l^n\right)k, \qquad (34d)$$

$$L_{u_m}^6 = g\left(u_m^n + \frac{1631}{55296}L_{u_m}^1 + \frac{175}{512}L_{u_m}^2 + \frac{575}{13824}L_{u_m}^3 + \frac{44275}{110592}L_{u_m}^4 + \frac{253}{4096}L_{u_m}^5, w_m^n + \frac{1631}{55296}L_{w_m}^1 \right.$$
$$\left. + \frac{175}{512}L_{w_m}^2 + \frac{575}{13824}L_{w_m}^3 + \frac{44275}{110592}L_{w_m}^4 + \frac{253}{4096}L_{w_m}^5, u_l^n\right)k. \qquad (34e)$$

Similarly, the delta is computed in the same fashion. It is important to mention as described in (34) that the coupled regime is fixed when computing the asset option and option sensitivities using the RKF method. Once $u_m^{n+1}$ and $w_m^{n+1}$ are known, $u_l^{n+1}$ and $w_l^{n+1}$ are computed from the latter. If (33c) does not hold, a new time step is computed from the previous one based on the relationship as follows [28]:

$$k_{new} = 0.9k_{old}(Tol/e_u)^{1/4}, \qquad \varepsilon > e_u. \qquad (35a)$$

However, if the condition in (33c) holds, then the numerical approximations for the optimal exercise boundary, asset option, and delta sensitivity for each regime are accepted and a new time step is also estimated which will be used in the next time level [28] as follows:

$$k_{new} = 0.9k_{old}(Tol/e_u)^{1/5}, \qquad \varepsilon \leq e_u. \qquad (35b)$$

## 3.3. Numerical Algorithm

Here, we present a numerical algorithm for approximating the optimal exercise boundary and asset and delta options based on the RKF methods with the coefficient entries of Cash and Karp [3]. The algorithm is described below.

---

**Algorithm.** Algorithm for the proposed method.

1. initialize $t = 0, h, k, T, A, B,$ and $Tol$    ▷ Arbitrary value of $k$ can be selected for a given $h$
2. initialize $s_{f(m)}^n, \boldsymbol{u}_m^n, \boldsymbol{w}_m^n$
3. **while** $t < T$
4.  **if** $t + k > T$
5.   $k = T - t$
6.  **endif**
7.  **while** true
8.   compute $s_{f(m)}^{n(i)}$, for $i = 1,2,3,4,5$    ▷ based on (25)
9.   compute $s_{f(m)}^{n+1}$
10.   **if** $s_{f(m)}^{n+1}$ is a real value, **break**    ▷ fix maximum value for $k$
11.   **else** $k = \phi k$    ▷ $0.1 \leq \phi \leq 0.5$ if the initial choice of $h^2 \leq k \leq h$
12.   **endif**
13.  **end while**
14.  compute $(\boldsymbol{f}_m^n)_u$ and $(\boldsymbol{f}_m^n)_w$
15.  compute $\boldsymbol{u}_m^{n+1}, \bar{\boldsymbol{u}}_m^{n+1}$, and $\boldsymbol{w}_m^{n+1}$    ▷ based on (22) and (23)
16.  compute $e_u = \max\limits_{1 \leq m \leq l} \left\| \bar{\boldsymbol{u}}_m^{n+1} - \boldsymbol{u}_m^{n+1} \right\|_\infty$
17.  **if** $e_u < Tol$
18.   set $\boldsymbol{u}_m^n = \boldsymbol{u}_m^{n+1}, \boldsymbol{w}_m^n = \boldsymbol{w}_m^{n+1}$, and $s_{f(m)}^n = s_{f(m)}^{n+1}$
19.   set $\delta_u = 0.9(Tol/e_u)^{1/4}$ and $k = \delta_u k$    ▷ based on (24)
20.   $t = t + k$
21.  **else**
22.   set $\delta_u = 0.9(Tol/e_u)^{1/5}$ and $k = \delta_u k$    ▷ based on (24)
23.  **endif**
24. **repeat**

---

## 4. Numerical Results

In this section, we consider the two- and four-regime examples using data from the existing works of literature. We experiment in a mesh with a uniform grid size and adaptive time step and compare our results with the existing methods. Furthermore, we chose $\phi = 0.5$ and $k = h^2$.

## 4.1. Two-Regimes example

Consider the example from the work of Khaliq and Liu [10] with the following data

$$Q = \begin{bmatrix} -6 & 6 \\ 9 & -9 \end{bmatrix}, \quad r = \begin{bmatrix} 0.10 \\ 0.05 \end{bmatrix}, \quad \sigma = \begin{bmatrix} 0.80 \\ 0.30 \end{bmatrix}, \quad \varepsilon = 10^{-6}, \quad T = 1, \quad K = 9 \qquad (36)$$

In our computation, we chose the interval $0 \leq x_m \leq 3$ with the grid size $h = 0.5, 0.025, 0.0125,$ and $0.01$. We labeled our method "CS-RKF" and compared our result with the MTree [18], MOL [4], and IMS1, and IMS2 [10]. We displayed the profiles of the asset and delta options and optimal exercise boundary for each regime in Fig. 1 and Table 1. In Table 2, we listed the total runtime in seconds with respect to $h$. Furthermore, the plot of the optimal time step selection at each time level was displayed in Fig. 2.

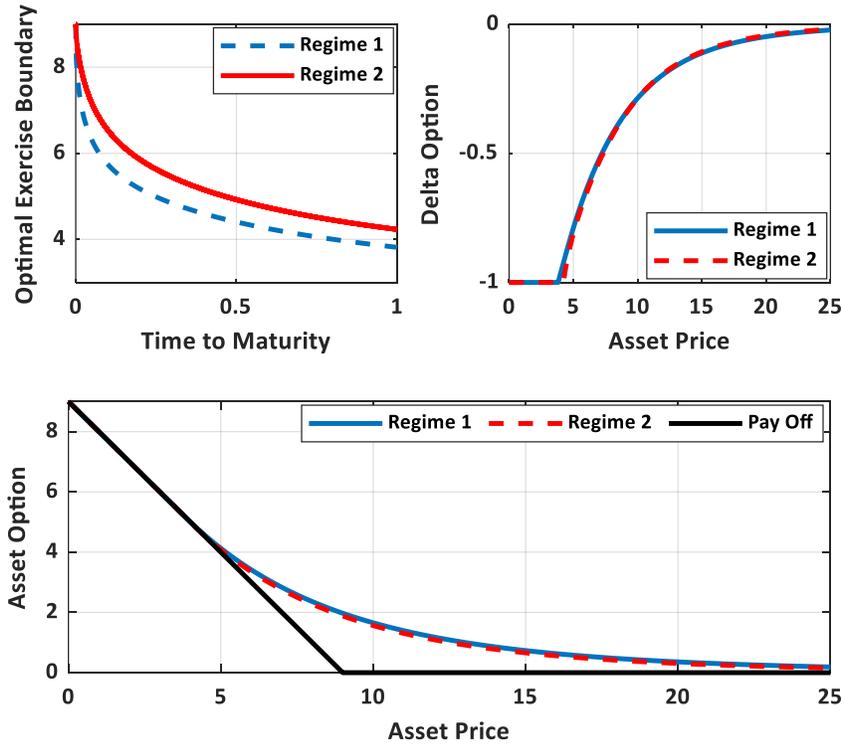

**Fig. 1.** Optimal exercise boundaries and asset and delta options for the two-regime case ($\tau = T, h = 0.0125$).

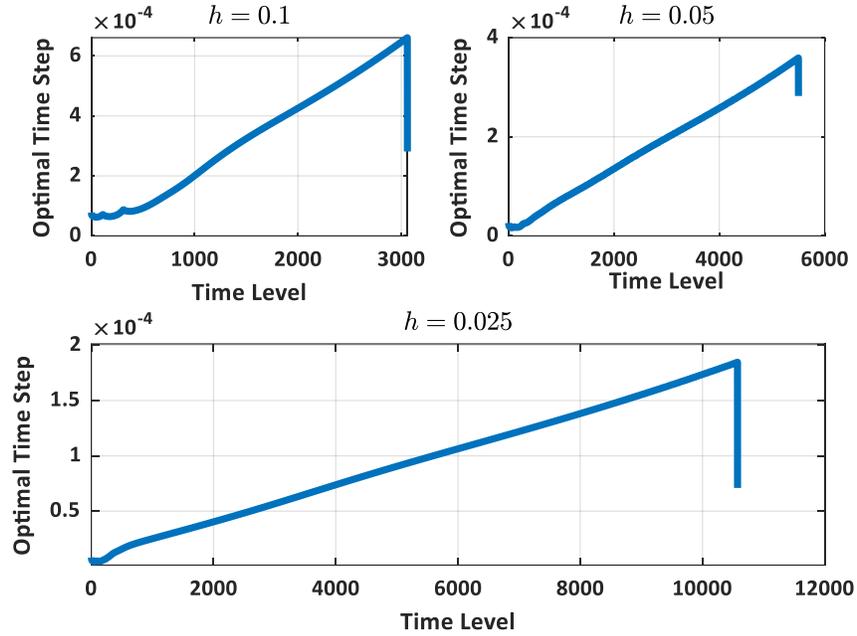

**Fig. 2.** Optimal time step selection per time level for the two-regime case ($\varepsilon = 10^{-6}$).

**Table 1.** Comparison of the American put option price for the two-regimes case.

| S | MTree | IMS1 | IMS2 | MOL | CS-RKF | | | |
|---|---|---|---|---|---|---|---|---|
| | | | | | $h = 0.05$ | 0.025 | 0.0125 | 0.01 |
| | | | | Regime 1 | | | | |
| 3.5 | 5.5000 | 5.5001 | 5.5001 | 5.5000 | 5.5000 | 5.5000 | 5.5000 | 5.5000 |
| 4.0 | 5.0066 | 5.0067 | 5.0066 | 5.0033 | 5.0033 | 5.0033 | 5.0033 | 5.0033 |
| 4.5 | 4.5432 | 4.5486 | 4.5482 | 4.5433 | 4.5440 | 4.5435 | 4.5434 | 4.5434 |
| 6.0 | 3.4144 | 3.4198 | 3.4184 | 3.4143 | 3.4142 | 3.4143 | 3.4143 | 3.4143 |
| 7.5 | 2.5844 | 2.5877 | 2.5867 | 2.5842 | 2.5853 | 2.5840 | 2.5842 | 2.5842 |
| 8.5 | 2.1560 | 2.1598 | 2.1574 | 2.1559 | 2.1556 | 2.1557 | 2.1559 | 2.1558 |
| 9.0 | 1.9722 | 1.9756 | 1.9731 | 1.9720 | 1.9725 | 1.9721 | 1.9720 | 1.9720 |
| 9.5 | 1.8058 | 1.8090 | 1.8064 | 1.8056 | 1.8064 | 1.8058 | 1.8056 | 1.8056 |
| 10.5 | 1.5186 | 1.5214 | 1.5187 | 1.5185 | 1.5193 | 1.5187 | 1.5185 | 1.5185 |
| 12.0 | 1.1803 | 1.1827 | 1.1799 | 1.1803 | 1.1802 | 1.1803 | 1.1803 | 1.1803 |
| | | | | Regime 2 | | | | |
| 3.5 | 5.5000 | 5.5012 | 5.5012 | 5.5000 | 5.5000 | 5.5000 | 5.5000 | 5.5000 |
| 4.0 | 5.0000 | 5.0016 | 5.0016 | 5.0000 | 5.0000 | 5.0000 | 5.0000 | 5.0000 |
| 4.5 | 4.5117 | 4.5194 | 4.5190 | 4.5119 | 4.5129 | 4.5122 | 4.5119 | 4.5119 |
| 6.0 | 3.3503 | 3.3565 | 3.3550 | 3.3507 | 3.3506 | 3.3505 | 3.3507 | 3.3507 |
| 7.5 | 2.5028 | 2.5078 | 2.5056 | 2.5033 | 2.5045 | 2.5032 | 2.5033 | 2.5033 |
| 8.5 | 2.0678 | 2.0722 | 2.0695 | 2.0683 | 2.0681 | 2.0682 | 2.0683 | 2.0683 |
| 9.0 | 1.8819 | 1.8860 | 1.8832 | 1.8825 | 1.8830 | 1.8825 | 1.8825 | 1.8825 |
| 9.5 | 1.7143 | 1.7181 | 1.7153 | 1.7149 | 1.7158 | 1.7150 | 1.7149 | 1.7148 |
| 10.5 | 1.4267 | 1.4301 | 1.4272 | 1.4273 | 1.4282 | 1.4275 | 1.4273 | 1.4273 |
| 12.0 | 1.0916 | 1.0945 | 1.0916 | 1.0923 | 1.0921 | 1.0924 | 1.0923 | 1.0923 |

**Table 2.** Total runtime for the two-regimes example.

| $h$ | Total runtime (s) |
|---|---|
| 0.1000 | 21 |
| 0.0500 | 41 |
| 0.0250 | 222 |
| 0.0100 | 2958 |

In Table 1, starting from $h = 0.025$, the values of the asset option in Regimes 1 and 2 are very close to the ones obtained from MTree [18] and MOL [4] methods. Hence, large step size is only required in our method to achieve an accurate solution. This is very useful and will be paramount when modeling a regime-switching problem beyond two regimes. Moreover, from Table 2, the total runtime for $h = 0.025$ is small even though we simultaneously compute the optimal exercise boundary and asset and delta options for each regime. Therefore, we can establish the advantage of our present method. Furthermore, we computed the convergent rate in space by using the fifth-order Runge-Kutta method based on the Cash and Karp coefficient in (33a) and (34a) with a constant step size $k = 2.5 \times 10^{-6}, T = 0.2$ and varying step sizes $h = 0.2, 0.1, 0.05, 0.025, 0.0125$. The result was listed in Table 3.

**Table 3.** Maximum errors and convergent rates in space of the asset option in Regime 1 ($k = 2.5 \times 10^{-6}$).

| | Asset Option | | Delta Option | |
|---|---|---|---|---|
| $h$ | maximum error | convergent rate | maximum error | convergent rate |
| 0.2 | ~ | ~ | ~ | ~ |
| 0.1 | $3.74 \times 10^{-1}$ | ~ | $6.72 \times 10^{-1}$ | ~ |
| 0.05 | $4.97 \times 10^{-2}$ | 2.914 | $8.20 \times 10^{-2}$ | 2.914 |
| 0.025 | $4.24 \times 10^{-3}$ | 3.551 | $6.96 \times 10^{-3}$ | 3.559 |
| 0.0125 | $2.21 \times 10^{-4}$ | 4.236 | $3.57 \times 10^{-4}$ | 4.282 |

From Table 3, the convergent rates of our method for both the asset and delta options are in close agreement with the theoretical convergent rate as the step sizes $h$ is reduced.

### 4.2. Four-Regimes example

Consider the four-regime switching model with the following data:

$$Q = \begin{bmatrix} -1 & 1/3 & 1/3 & 1/3 \\ 1/3 & -1 & 1/3 & 1/3 \\ 1/3 & 1/3 & -1 & 1/3 \\ 1/3 & 1/3 & 1/3 & -1 \end{bmatrix}, \quad r = \begin{bmatrix} 0.02 \\ 0.10 \\ 0.06 \\ 0.15 \end{bmatrix}, \quad \sigma = \begin{bmatrix} 0.90 \\ 0.50 \\ 0.70 \\ 0.20 \end{bmatrix}, \quad \varepsilon = 10^{-6}. \tag{37}$$

Here, we chose the interval $0 \leq x_m \leq 3$, used (17b) with four grid points and consider a step size $h = 0.05, 0.025, 0.0125,$ and $0.01$.

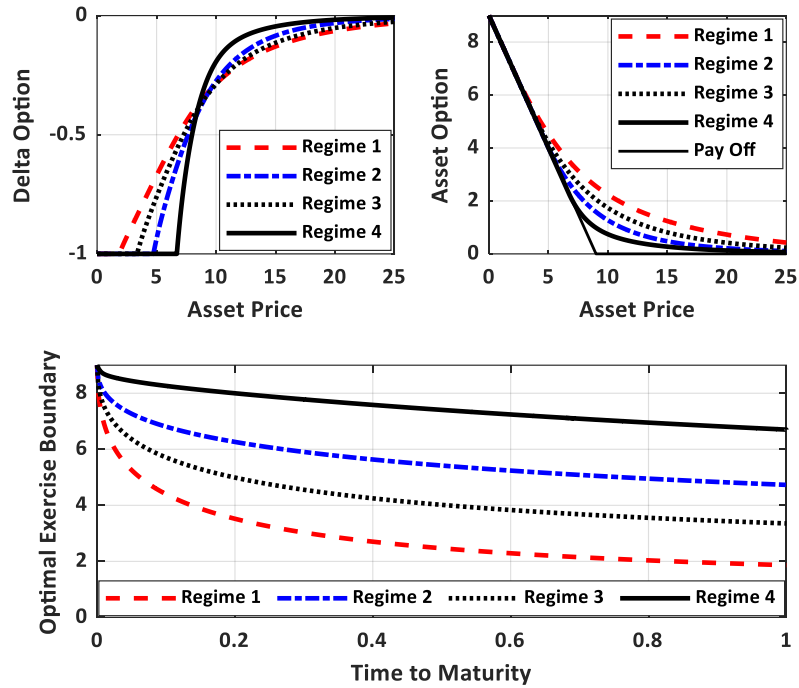

**Fig. 3.** Optimal exercise boundaries and the options for the four-regime case ($\tau = T, h = 0.025$).

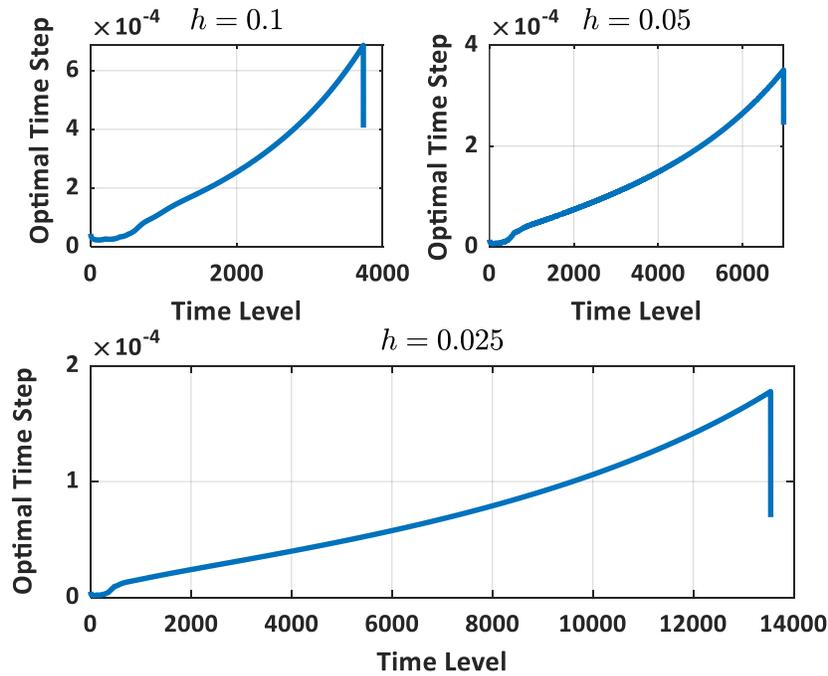

**Fig. 4.** Optimal time step selection per time level for the four-regime case ($\varepsilon = 10^{-6}$).

The profiles of the options and optimal exercise boundary for each regime were listed and displayed in Figs. 3 and 4 and Tables 4 and 5. From Table 4, the results obtained from our method are very close to those obtained from the MTree [18] and RBF-FD [16] methods. Moreover, the total runtime for the four-regimes example in Table 5 indicates that our method is very fast in computation.

**Table 4.** Comparison of American put options price for the four-regimes example.

| | MTree | | | | RBF-FD | | | |
|---|---|---|---|---|---|---|---|---|
| S | Reg 1 | Reg 2 | Reg 3 | Reg 4 | Reg 1 | Reg 2 | Reg 3 | Reg 4 |
| 7.5. | 3.1433 | 2.2319 | 2.6746 | 1.6574 | 3.1424 | 2.2320 | 2.6744 | 1.6576 |
| 9.0 | 2.5576 | 1.5834 | 2.0568 | 0.9855 | 2.5564 | 1.5835 | 2.0566 | 0.9857 |
| 10.5 | 2.1064 | 1.1417 | 1.6014 | 0.6533 | 2.1052 | 1.1415 | 1.6013 | 0.6554 |
| 12.0 | 1.7545 | 0.8377 | 1.2625 | 0.4708 | 1.7527 | 0.8377 | 1.2625 | 0.4708 |
| | ff-expl | | | | | | | |
| S | Reg 1 | | Reg 2 | | Reg 3 | | | Reg 4 |
| 7.5 | 3.1421 | | 2.2313 | | 2.6739 | | | 1.6573 |
| 9.0 | 2.5563 | | 1.5827 | | 2.0559 | | | 0.9850 |
| 10.5 | 2.1047 | | 1.1406 | | 1.6004 | | | 0.6546 |
| 12.0 | 1.7524 | | 0.8368 | | 1.2614 | | | 0.4700 |
| | CS-RKF | | | | | | | |
| | $h = 0.05$ | | | | 0.025 | | | |
| 7.5 | 3.1425 | 2.2337 | 2.6752 | 1.6631 | 3.1424 | 2.2321 | 2.6748 | 1.6589 |
| 9.0 | 2.5563 | 1.5847 | 2.0581 | 0.9874 | 2.5549 | 1.5838 | 2.0571 | 0.9862 |
| 10.5 | 2.1029 | 1.1419 | 1.6022 | 0.6566 | 2.1016 | 1.1414 | 1.6015 | 0.6553 |
| 12.0 | 1.7477 | 0.8392 | 1.2637 | 0.4714 | 1.7471 | 0.8376 | 1.2620 | 0.4708 |
| | 0.0125 | | | | 0.01 | | | |
| 7.5 | 3.1418 | 2.2319 | 2.6746 | 1.6675 | 3.1418 | 2.2319 | 2.6746 | 1.6577 |
| 9.0 | 2.5548 | 1.5835 | 2.0567 | 0.9858 | 2.5549 | 1.5835 | 2.0567 | 0.9858 |
| 10.5 | 2.1015 | 1.1414 | 1.6013 | 0.6553 | 2.1015 | 1.1413 | 1.6012 | 0.6553 |
| 12.0 | 1.7467 | 0.8375 | 1.2620 | 0.4706 | 1.7468 | 0.8374 | 1.2621 | 0.4706 |

**Table 5.** Total runtime for the four-regimes example.

| $h$ | Total runtime (s) |
|---|---|
| 0.1000 | 41 |
| 0.0500 | 91 |
| 0.0250 | 594 |

### 4.3. Computing the Gamma Option

The gamma option is one of the important parameters for hedging options. Hence, computing its value values with high order accuracy is paramount and ideal. Some of the previous literature approximated the gamma option using the numerical solution of the asset option for each regime. This procedure, sometimes, results in an inaccurate and spurious solution, especially beyond two-regime examples.

To approximate the gamma option with high order accuracy and avoid obtaining a spurious solution, we further introduce another nonlinear PDE in (4) for approximating the gamma option simultaneously with the asset and delta options and the optimal exercise boundary for each regime as follows:

$$\frac{\partial Y_m}{\partial \tau} - \frac{1}{2}\sigma^2_m \frac{\partial^2 Y_m}{\partial x_m^2} - \xi_m^\tau \frac{\partial^2 W_m}{\partial x_m^2} + (r_m - q_{mm})Y_m - \sum_{l \neq m} q_{ml} Y_l = 0, \quad x_m > 0, \quad (38a)$$

with the initial and boundary conditions given as follows:

$$Y_m(x_m, \tau) = -s_{f(m)} e^{x_m}, \quad x_m < 0; \quad (38b)$$

$$Y_m(x_m, 0) = 0, \quad x_m \geq 0, \quad Y_m(0, \tau) = -s_{f(m)}, \quad Y_m(\infty, \tau) = 0. \quad (38c)$$

Following the procedures in (29), (33), and (34) for the numerical approximation of (38), the plot profiles of the gamma option for the two and four-regimes examples were presented in Fig. 5.

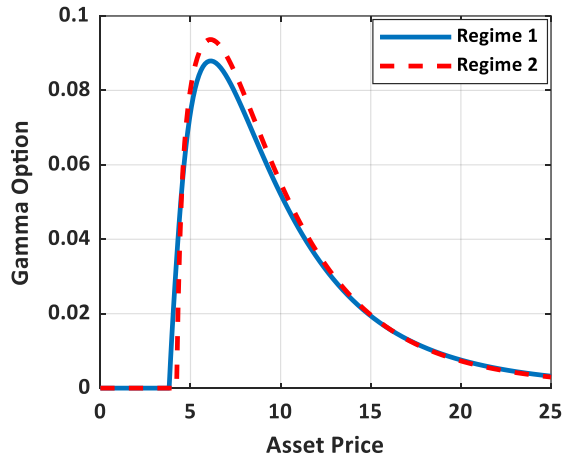

**Fig. 5a.** Gamma option for the two-regimes case ($\tau = T, h = 0.02$).

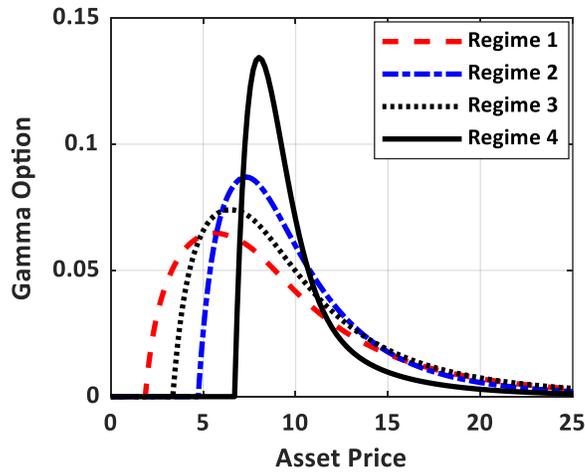

**Fig. 5b.** Gamma option for the four-regimes case ($\tau = T, h = 0.02$).

**Remark 2.** It is important to mention that other Greeks like speed, theta, delta decay, and color options can be simultaneously computed from our method. However, in this work, we intend to focus on the asset, delta, and gamma options using the two- and four-regimes examples as a case study.

## 5. Conclusion

We have proposed an explicit and high-order numerical method for solving the regime-switching model. We first recast the free boundary problem to a system of coupled nonlinear partial differential equations and further introduce a transformation based on square root function with fixed free boundary from which a high order analytical approximation is obtained for computing the derivative of the optimal exercise boundary in each regime. We then employ a high order compact finite difference scheme and RKF adaptive time integration for the discretization and approximation of the optimal exercise boundary, options value, and option Greeks in the set of coupled ODEs and each regime. The convergent rate is further computed and is in good agreement with the theoretical convergent rate as $h$ decreases. By further comparing our results with the existing methods, we then validate that our method performs better in terms of accuracy and computational speed.

## Data Availability Statement

The datasets during and/or analyzed during the current study available from the corresponding author on reasonable request.

## Funding

No funding was received from any resource.

## Conflict of interest

The author declares that he has no conflict of interest.